\documentclass{elsart}
\usepackage{amssymb,epsf,mathptm,amsmath}
\begin{document}
\begin{frontmatter}
\title{Sandpile avalanche dynamics on scale-free networks}
\author{D.-S.~Lee, K.-I.~Goh, B. Kahng and D. Kim}
\address{
School of Physics and Center for Theoretical Physics, Seoul
National University NS50, Seoul 151-747, Korea}
%\date{\today}
\begin{abstract}
Avalanche dynamics is an indispensable feature of complex
systems. Here we study the self-organized critical dynamics of
avalanches on scale-free networks with degree exponent $\gamma$
through the Bak-Tang-Wiesenfeld (BTW) sandpile model. The
threshold height of a node $i$ is set as $k_i^{1-\eta}$ with
$0\leq\eta<1$, where $k_i$ is the degree of node $i$. Using the
branching process approach, we obtain the avalanche size and the
duration distribution of sand toppling, which follow power-laws
with exponents $\tau$ and $\delta$, respectively. They are given
as $\tau=(\gamma-2 \eta)/(\gamma-1-\eta)$ and
$\delta=(\gamma-1-\eta)/(\gamma-2)$ for $\gamma<3-\eta$, $3/2$ and
$2$ for $\gamma>3-\eta$, respectively.
The power-law distributions are modified by a logarithmic correction 
at $\gamma=3-\eta$.
\end{abstract}
\begin{keyword}
avalanche \sep
scale-free network \sep
branching process
\PACS
89.70.+c \sep
89.75.-k \sep
05.10.-a
\end{keyword}
\end{frontmatter}

\section{Introduction}
Frequently, complex systems in nature as well as in human society
suffer  massive catastrophes triggered from only a small fraction
of their constituents. Unexpected epidemic spread of diseases and
the power outage in the  eastern US of the last year are the
examples of such avalanche phenomena. Such a cascading dynamics is 
not always harmful to us. The information cascades making popular
hits of books, movies, and albums are good to writers, actors, and
singers, respectively. Thus it is interesting to understand and
predict how those cascades propagate in complex system. Recently,
the network approach, by which a system is viewed as a network
consisting of nodes representing its constituents and links
interactions between them, simplifies complicated details of
complex systems. Such a simplification unveils a hidden order such
as scale-free behavior in the degree distribution. Here degree is
the number of links connected to a certain node. The Internet at
the autonomous system level, the World-Wide Web, social
acquaintance networks, biological networks, and other many complex
networks exhibit power-law degree distributions, $p_d(k)\sim
k^{-\gamma}$. The networks following such power-law degree
distributions are called scale-free (SF) networks~\cite{ba99},
where non-negligible fractions of hubs, the nodes with
extraneously large degrees, exist.

In this paper, we investigate the avalanche dynamics on such SF
networks through the Bak-Tang-Wiesenfeld (BTW) sandpile model
~\cite{btw}, a prototypical model exhibiting self-organized
criticality (SOC). The study of sandpile dynamics has been carried
out mostly on regular lattices in the Euclidean space. In the
stationary state, which can be reached without tuning a parameter,
the system exhibit scale-invariant features in the power-law form of the
avalanche size distribution $p_a(s)$ and the duration or lifetime
distribution $\ell(t)$ as
\begin{equation}
p_a(s) \sim s^{-\tau} \quad{\rm and}\quad   \ell(t)\sim t^{-\delta}.
\label{distr}
\end{equation}
Recently, Bonabeau has studied the sandpile dynamics on the
Erd\H{o}s-R\'enyi (ER) random networks~\cite{bonabeau} and found
that the avalanche size distribution follows a power law with the
exponent $\tau \simeq 1.5$, consistent with the mean-field
solution~\cite{alstrom}. Recently, Lise and Paczuski
~\cite{paczuski} studied the Olami-Feder-Christensen
model~\cite{ofc} on regular ER networks, where degree of each node
is uniform but connections are random. They found the exponent to
be $\tau\approx1.65$. However, when degree of each node is not
uniform, they found no criticality in the avalanche size
distribution. Note that they assumed that the threshold of each
node is uniform, whereas degree is not. Here we study the BTW
sandpile model on SF networks, where the threshold $z_i$ of the
node $i$ is given as $k_i^{1-\eta}$ with $k_i$ the degree of $i$
and $0\leq \eta <1$. We find that the exponents for the avalanche
size and the duration distribution depend on the degree exponent
$\gamma$ as $\tau=(\gamma-2 \eta)/(\gamma-1-\eta)$ and
$\delta=(\gamma-1-\eta)/(\gamma-2)$ for $\gamma<3-\eta$ while, for
$\gamma>3-\eta$,  they show the same behaviors as the conventional
mean-field solutions as observed for the ER random networks.

\section{Sandpile model}
We present the dynamic rule of the BTW sandpile model on a given
network.
\begin{enumerate}
\item Each node $i$ is given a prescribed threshold $z_i$ $(\leq k_i)$.  
The smallest integer not smaller than $z_i$ is denoted as 
$\lceil z_i \rceil$ ($\lceil z_i \rceil\le k_i$).
\item At each time step, a grain is added at a randomly chosen node $i$.
The integer-valued height of the node $i$, $h_i$, increases by $1$.
\item If the height at the node $i$ reaches or exceeds 
$z_i$,  then it becomes unstable and the $\lceil z_i\rceil$ grains
at the node topple to its $\lceil z_i\rceil$ 
randomly chosen  adjacent nodes among $k_i$
ones;\\
$h_i \rightarrow h_i-\lceil z_i\rceil$, 
and $h_j=h_j+1$ for all nodes $j$ which are chosen. 
\item If this toppling causes any of the adjacent nodes receiving grains
to be unstable, subsequent topplings follow on those nodes
in parallel until there is no unstable node left.
This process defines an avalanche.
\item Repeat (2)--(4).
\end{enumerate}
Here the threshold $z_i$ of node $i$ is given as
\begin{equation}
z_i=k_i^{1-\eta} ~(0\leq \eta<1),
\label{eq:threshold}
\end{equation}
which is a generalization of $z_i=k_i$ previously investigated in
Ref.~\cite{goh03}. 
We concentrate on the distributions of (i) the
avalanche area $A$, $i.e.,$ the number of distinct nodes involving
in a given avalanche, (ii) the avalanche size $S$, $i.e.,$ the
number of toppling events in a given avalanche, and (iii) the
duration $T$ of a given avalanche.

\section{Branching process approach}
The mapping of each avalanche to a tree provides a useful way of
understanding the statistics of avalanche dynamics analytically.
For each avalanche event, one can draw a corresponding tree: The
node where an avalanche is triggered corresponds to the originator
of the tree and the following nodes to descendants. In the tree
structure, a descendant born at time $t$ is located away from the
originator by distance $t$ along the shortest pathway. The tree
stops to grow when no further avalanche proceeds. Then the
ensemble of avalanches can be identified with that of trees grown
through the branching process. In this mapping, the avalanche
duration $T$ is equal to the lifetime of the tree minus one, and
the avalanche size $S$ differs from the tree size only by the
number of boundary nodes of the tree, which is relatively small
when the overall tree size is very large. If one assumes that
branching events at different nodes occur independently  and that
there is no loop in the tree, the tree size and lifetime
distribution can be obtained analytically~\cite{harris,otter}.
Those distributions are expected to share the same asymptotic
behaviors with the avalanche size and duration distribution,
respectively, due to the near-equivalence between an avalanche and
its corresponding tree in their scales as mentioned above.

In the branching process describing an avalanche, 
after initial branching into $k$ descendants with probability $q_0(k)$, 
successive branchings are assumed to occur independently with 
probability $q(k)$.  $q_0(k)$ and $q(k)$ may be different in general, but  
the statistics of the overall size and duration of an avalanche is 
determined dominantly by $q(k)$. We checked also numerically the case 
where a new grain is added to a node with the probability proportional to 
the degree of that node, which gives different $q_0(k)$ from 
that in the case where a new grain is added randomly, 
and found that the nature of the avalanche dynamics is the same in both cases.
Thus, for simplicity, we consider  the branching process 
where every branching occur with probability $q(k)$. 
For the BTW model in the Euclidean space, where the threshold $z_i$ of
node $i$ is equal to its degree $k_i$, $q(k)$ has a finite cut-off
such that $q(k)=0$ for $k> z_i={\rm const}$, because the degree of
each node is uniform and finite. Consequently, the exponents of
the avalanche size and the duration distributions in
Eq.~(\ref{distr}) come out to be the so-called mean-field values; 
$\tau=3/2$ and $\delta=2$~\cite{harris,otter}. These results are known to hold
for the BTW model on regular lattices with dimensions larger than
$4$~\cite{alstrom}. Note that when dimension is smaller than 4,
the branching process approach cannot be applied, so that the
values of the exponents $\tau$ and $\delta$ would not be trivial.

In SF networks, avalanches usually do not form loops, generating
tree-structures: According to the numerical simulations of the BTW
model for the case of $z_i=k_i$ on SF networks~\cite{goh03}, the statistics of
the two quantities $A$ and $S$ are nearly equal when they are
large:  For example, the maximum area and size ($A_{\rm max}$,
$S_{\rm max}$) among avalanches are (5127, 5128), (12058, 12059)
and (19692, 19692) for scale-free networks with $\gamma=2.01$,
$3.0,$ and $\infty$, respectively. The fact that $A$ and $S$ are
almost the same implies that the avalanche structure can be
treated as a tree. From now on, we shall not distinguish $A$ and
$S$, denoted by $s$. Thus it is valid to use the branching process
approach to understand the avalanche dynamics on SF networks.

We study the BTW model on SF networks with the degree exponent
$\gamma$ and the threshold given as Eq.~(\ref{eq:threshold}). The
branching probability $q(k)$ consists of two factors, that is,
$q(k)=q_1 (k) q_2 (k)$, where $q_1(k)$ is the probability that the
threshold $z_i$ of node $i$ is in the range $k-1 <z_i \le k$ and $q_2(k)$ is
the probability that the total number of grains at the node
reaches or exceeds the threshold. 
If $z_i=f(k_i)$ with $f(x)$ a monotonic increasing function of $x$ satisfying 
$f(x)\leq x$ for all $x\geq 1$, 
the condition of $k-1 <z_i \le k$ implies that $q_1(k)$ is nothing but
the probability that a node $i$ connected to the one end of a
randomly chosen edge has its degree $k_i$ in the region
$(f^{-1}(k-1),f^{-1}(k)]$, and thus
$q_1(k)=\sum_{k'=\lfloor f^{-1}(k-1)\rfloor+1} ^{\lfloor
f^{-1}(k)\rfloor} k'p_d(k')/\langle k\rangle$, 
where $\lfloor x \rfloor$ is the largest integer not larger than   $x$. 
Notice that $\sum_{k=1}^\infty q_1(k)=1$ and 
$q_1(k)\sim k^{(1-\gamma+\eta)/(1-\eta)}$ for large $k$ if 
$f(x)\simeq x^{1-\eta}$ $(0\leq \eta<1)$ for large $x$.
$q_2(k)$ is the probability that the
node $i$ has height $k-1$ at the moment of receiving a grain from
one of its neighbors. We have checked numerically that a typical
height of node is absent, so that 
all possible $k$ values $0, 1,\ldots, k-1$ are equally likely~\cite{jkps}.
Thus we set $q_2(k)=1/k$.
As a result, the branching probability $q(k)$ for large $k$ is
given asymptotically as
\begin{equation}
q(k)=\frac{1}{k}q_1(k)\sim k^{-\gamma'} \quad
\left(
\gamma'={\gamma-2\eta\over 1-\eta}
\right).
\label{eq:qk}
\end{equation}
When $z_i=k_i$ or $\eta=0$, $\gamma'$ is reduced to
$\gamma$. Since we are interested in the case of $\gamma > 2$ and
$0 \le \eta < 1$, $\gamma' > 0$.

Using the independence of the branchings from different parent-nodes,
one can derive the following self-consistent relation for the
tree size distribution $p(s)$ as~\cite{harris,otter}
\begin{eqnarray}
p(s)&=&q(0)\delta_{s,1}+\sum_{k=1}^\infty q(k) \sum_{s_1=1}^\infty
\sum_{s_2=1}^\infty \cdots \sum_{s_k=1}^\infty
p(s_1) p(s_2) \ldots p(s_k)
\delta_{\sum_{i=1}^k s_i, s-1}.
\label{eq:recur}
\end{eqnarray}
This relation can be written in a more compact form
by introducing the generating functions,
$\mathcal{P}(y)=\sum_{s=1}^\infty p(s) y^s$ and
$\mathcal{Q}(\omega)=\sum_{k=0}^\infty q(k) \omega^k$ as
\begin{equation}
\mathcal{P}(y) = y \ \mathcal{Q}(\mathcal{P}(y)).
\label{eq:sc}
\end{equation}
Then $\omega=\mathcal{P}(y)$ is obtained by
inverting $y=\mathcal{P}^{-1}(\omega)=\omega/\mathcal{Q}(\omega)$.

The average size $\langle s \rangle$ of a finite tree can be
obtained easily from the generating functions.
\begin{equation}
\langle s \rangle = \sum_{s=1}^{\rm finite} s
p(s)=\mathcal{P'}(1),
\end{equation}
where $\mathcal{P'}(y)=d\mathcal{P}(y)/dy$. 
Using the relation, Eq.~(\ref{eq:sc}), we obtain
\begin{equation}
\langle s \rangle=
\mathcal{P'}(1)=\frac{\mathcal{Q}(\mathcal{P}(1))}{1-\mathcal{Q'}(\mathcal
{P}(1))},
\label{eq:meansize}
\end{equation}
where again $\mathcal{Q'}(\omega)=d\mathcal{Q}(\omega)/d\omega$.

The distribution of duration, $i.e.$, the lifetime of the tree can
be evaluated similarly~\cite{harris,otter}. Let $r(t)$ be the
probability that a branching process stops at or prior to time
$t$. Then following the similar steps leading to
Eq.~(\ref{eq:recur}), $i.e.$, $r(t)=\sum_{k=0}^{\infty}q_k
[r(t-1)]^k$, one has
\begin{equation}
r(t)=\mathcal{Q}(r(t-1)).
\label{eq:life}
\end{equation}
For large $t$, $r(t)$ comes close to $1$.
One can obtain $\omega=r(t-1)$ by solving
$d\omega /dt \simeq r(t)-r(t-1)=Q(\omega)-\omega$.
Then the lifetime distribution $\ell(t)$ is obtained through
$\ell(t)=r(t)-r(t-1)\simeq d\omega/dt$.

\section{Avalanche size and duration distribution}

The growth of a tree depends on the average number of branches defined
as
\begin{equation}
C=\sum_{k=1}^\infty k q(k).
\end{equation}
When $C>1$ ($C<1$), a  tree can (cannot) grow infinitely in a
probabilistic sense. Thus the case of $C=1$ is a critical point
for the growth of a tree. One can see that 
for any branching process with $q(k)=(1/k) q_1(k)$ ($k\geq 1$) and 
$\sum_{k=1}^\infty q_1(k)=1$, 
the average number of branches $C$ is always $1$, independent of
detailed structural properties of networks. 
Therefore our assumption  $q_2(k)=1/k$ 
corresponds to the condition for 
the self-organized criticality (SOC) of the sandpile model.

The inverse function $\mathcal{P}^{-1}(\omega)$ satisfies
$\mathcal{P}^{-1}(1)=1$. When $C=1$, the first-order derivative
$\partial \mathcal{P}^{-1}(\omega)/\partial \omega$ at $\omega=1$
is zero and thus $\mathcal{P}(y)$ becomes singular at $y=1$. 
$\mathcal{P}(y)$ is expanded around $y=1$ as $\mathcal{P}(y)\simeq
1-b (1-y)^\phi$ with constant $b$ and $0<\phi<1$. Then the
asymptotic behavior of the avalanche size distribution $p(s)$ for
large $s$ is given by $p(s)\sim s^{-\phi-1}$, because
if a series $\sum_{s=0}^\infty a_s y^s$ with the radius of convergence 
$1$ has the asymptotic behavior 
\begin{equation}
\sum_{s=0}^\infty a_s y^s\sim (1-y)^\phi ~~{\rm as}~~ y\to 1, ~~ {\rm then}~~ a_s 
\sim s^{-\phi-1}~~~ {\rm as}~~~ s\to\infty. \label{eq:al_sing}
\end{equation}

The functional form of the branching probability $q(k)$ determines
the singularity of $\mathcal{P}(y)$. To illustrate this, we first
consider a simple case that
\begin{equation}
q(k)=\left\{
\begin{tabular}{ll}
$1-a$ & \phantom{a}($k=0$), \\
$a$ & \phantom{a}($k=2$), \\
$0$ & \phantom{a}(${\rm otherwise}$),
\end{tabular}\right.
\label{eq:example_qk}
\end{equation}
where $0<a<1$. 
Then the average number of branches $C=\sum kq(k)=2a$ and the
generating function $\mathcal{Q}(\omega)=\sum_{k=0}^{\infty} q(k)
\omega^k = 1-a+a\omega^2$. Using the relations of
$y=\omega/\mathcal{Q}(\omega)$ and $\omega=\mathcal{P}(y)$, it is
obtained that
\begin{equation}
\mathcal{P}(y)=\frac{1-\sqrt{1-4a(1-a)y^2}}{2ay}.
\label{eq:example_py}
\end{equation}
The value of $\mathcal{P}(1)=\sum_{s=1}^{\rm finite}p(s)$ 
is given as 
\begin{equation}
\mathcal{P}(1)=\frac{1-|1-2a|}{2a}=\left\{
\begin{tabular}{ll}
$1$ & \phantom{a} {\rm for}~~~ $0<a\le {1 \over 2}$~~~($C \le 1$), \\
$\frac{1-a}{a}$ & \phantom{a} {\rm for}~~~ ${1\over 2} < a <
1$~~~($C
> 1$),
\end{tabular}\right.
\label{eq:example_p(1)}
\end{equation}
which means that when $1/2<a<1$ ($C>1$), 
a tree can grow infinitely with probability $1-\mathcal{P}(1)=(2a-1)/a$, 
and the critical point is located at $a_c=1/2$. 
Near $y=1$,
$\mathcal{P}(y)\approx 1-\sqrt{2(1-y)}$ from
Eq.~(\ref{eq:example_py}), leading to $\phi=1/2$. Then, the
avalanche size distribution $p(s)$ behaves as $p(s)\sim s^{-3/2}$.
On the other hand, 
using Eq.~(\ref{eq:meansize}), 
\begin{equation}
\langle s \rangle =\mathcal{P}'(1)=\left\{
\begin{tabular}{ll}
$\frac{1}{2(a_c-a)}$ & \phantom{a}~~ ($a < a_c$), \\
$\frac{1-a}{2a(a-a_c)}$ & \phantom{a}~~ $(a > a_c)$.
\end{tabular}\right.
\label{eq:example_smean}
\end{equation}

Even for the case that $q(k)$ has a finite cut-off larger than $2$ or 
decays exponentially,  the above result holds.
This is the conventional mean-field solution for the avalanche size
distribution~\cite{alstrom,harris,otter} and has been shown to hold
for the BTW model on the ER random networks~\cite{bonabeau}.

When $q(k)$ decays slowly as in Eq.~(\ref{eq:qk}), however, its
generating function $\mathcal{Q}(\omega)$ is singular at
$\omega=1$. For $q(k)$ in Eq.~(\ref{eq:qk}), the expansion of
$\mathcal{Q}(\omega)$ around $\omega=1$ is given as
\begin{equation}
\mathcal{Q}(\omega)\simeq 1-(1-\omega)+
\left\{
\begin{tabular}{ll}
$A_1 \, (1-\omega)^{\gamma'-1}$ & \phantom{a}($2<\gamma <\gamma_c$), \\
$-A_2 \, (1-\omega)^2\ln (1-\omega)$ & \phantom{a}($\gamma =\gamma_c$), \\
$A_3 \, (1-\omega)^{2}$ & \phantom{a}($\gamma >\gamma_c$),
\end{tabular}\right.
\label{eq:q_expand}
\end{equation}
where $A_i$'s are constants, $\gamma'$ is given in Eq.~(\ref{eq:qk}),
and $\gamma_c=3-\eta$.
The derivation of the logarithmic correction for the case of $\gamma=\gamma_c$  
can be found in
~\cite{robin}. Note that the singular term
$(1-\omega)^{\gamma'-1}$ is the second leading term of
$1-\mathcal{Q}(\omega)$ for $\gamma<\gamma_c$. Using the relation
$\mathcal{P}^{-1}(\omega)=\omega/\mathcal{Q}(\omega)$ in
Eq.~(\ref{eq:sc}), the behavior of $\mathcal{P}(y)$ around $y=1$
is obtained for each region of $\gamma$ from
Eq.~(\ref{eq:q_expand}), and in turn, using
Eq.~(\ref{eq:al_sing}), $p(s)$ for $s\to\infty$. 
We find that 
\begin{equation}
p(s)\sim \left\{
\begin{array}{ll}
s^{-(\gamma-2\eta)/(\gamma-1-\eta)} & (2< \gamma <\gamma_c), \\[2.5mm]
s^{-3/2}(\ln s)^{-1/2} & (\gamma=\gamma_c),\\[2.5mm]
s^{-3/2} & (\gamma>\gamma_c).
\end{array}
\right.
\label{eq:P(s)}
\end{equation}
Thus, the exponent $\tau$ is given as
$\tau=(\gamma-2\eta)/(\gamma-1-\eta)$ for $2<\gamma < \gamma_c$
and $\tau=3/2$ for $\gamma \ge \gamma_c$.

Also obtained is $r(t)$ from Eq.~(\ref{eq:q_expand}) by using
Eq.~(\ref{eq:life}). The duration distribution $\ell(t)$, which is
the derivative of $r(t)$, is found  to be
\begin{equation}
\ell(t)\sim \left\{
\begin{array}{ll}
t^{-(\gamma-1-\eta)/(\gamma-2)} & (2< \gamma <\gamma_c), \\[2.5mm]
t^{-2}(\ln t)^{-1} & (\gamma=\gamma_c), \\[2.5mm]
t^{-2}  & (\gamma>\gamma_c).
\end{array}
\right.
\label{eq:D(t)}
\end{equation}
That is, the exponent $\delta$ is given as $\delta=(\gamma-1-\eta)/(\gamma-2)$
for $2<\gamma<\gamma_c$ and $\delta=2$ for $\gamma\geq \gamma_c$.

\section{Conclusion}
We have studied the BTW sandpile model on SF networks with 
the degree exponent $\gamma$ to
understand the avalanche dynamics in complex systems. The main
results are the avalanche size and duration distribution. The
exponents $\tau$ and $\delta$ increase with increasing $\gamma$,
implying that the hubs play a role of reservoir, that is,
sustain large amount of grains to make the SF network resilient
under avalanche dynamics. 
This is reminiscent of the structural
resilience of the SF network under random removal of nodes for
$\gamma\leq 3$~\cite{bara,cohen1,newman1}. We also checked the
case where the threshold $z_i$ contains noise in the way that
$z_i=\zeta_i k_i$ with $\zeta_i$ being distributed uniformly in
[0,1].  We find that
such a variation does not change the nature of the avalanche
dynamics. However, when the threshold is given in terms of a
quantity other than degree, e.g., load, the corresponding
avalanche dynamics has no reason to follow the same statistics as
studied in this paper, which remains further works.
\\
This work is supported by the KOSEF Grant No. R14-2002-059-01000-0
in the ABRL program.

\end{document}